\documentclass[conference]{IEEEtran}
\ifCLASSINFOpdf
\else
\fi

\usepackage{graphicx}
\usepackage{adjustbox}
\usepackage{amsmath}

\begin{document}
\title{Risk Analysis Study of Fully \\ Autonomous Vehicle}

\author{
    \IEEEauthorblockN{Chuka Oham}
    \IEEEauthorblockA  {
       School of Computer Science and Engineering \\
        University of New South Wales\\ New South Wales, Australia \\
        CSIRO, Data 61 %E-mail: chuka.oham@csiro.au\\
        %Country of Z
    }
    \and
    \IEEEauthorblockN{Raja Jurdak}
    \IEEEauthorblockA  {
        CSIRO, Data 61\\
        Brisbane, Queensland \\ Australia\\
        IEEE Senior Member}
          \and
    \IEEEauthorblockN{Sanjay Jha}
    \IEEEauthorblockA  {
        School of Computer Science and Engineering \\
        University of New South Wales\\ New South Wales, Australia \\ IEEE Senior Member
    }
      }
\maketitle

\begin{abstract}
Fully autonomous vehicles are emerging vehicular technologies that have gained significant attention in today’s research endeavours. Even though it promises to optimize road safety, the proliferation of wireless and sensor technologies makes it susceptible to cyber threats thus dawdling its adoption. The identification of threats and design of apposite security solutions is therefore pertinent to expedite its adoption. In this paper, we analyse the security risks of the communication infrastructure for the fully autonomous vehicle using a subset of the TVRA methodology by ETSI. We described the model of communication infrastructure. This model clarifies the potential communication possibilities of the vehicle. Then we defined the security objectives and identified threats. Furthermore, we classified risks and propose countermeasures to facilitate the design of security solutions. We find that all identified high impact threats emanates from a particular source and required encryption mechanisms as countermeasures. Finally, we discovered that all threats due to an interaction with humans are of serious consequences.   
\end{abstract}


% \fi
\IEEEpeerreviewmaketitle

\section{Introduction}
Autonomous vehicles are vehicular technologies with the ability to make independent driving decisions. They are cooperative systems which aim at the optimization of the road transport systems and road safety [2] and are based on the following vehicular interactions; vehicle-to-vehicles (v2v), vehicle-to-infrastructure (v2i) and vehicle-to-other connecting technologies (v2x). Research into autonomous vehicle technology has received significant attention from academia and automotive industry and also received support from the government due to its potential benefits [1, 3]. \\
NHTSA [4] defines 5 levels of autonomy to reflect the degree to which a driver is required to control and monitor road operations. Levels 0-2 require the driver to absolutely monitor the roadway situation. Level 3 requires the driver to maintain alertness to know when to take back control of the vehicle while in level 4, the driver cedes absolute control of critical driving functions and road monitoring. \\
The generation of valuable data and exposure to external communications makes it susceptible to cyber threats with life threatening outcomes upon the successful execution of a malicious threat. Therefore, to expedite the adoption of this technology, it must be able to maximally guarantee the security of the driver and passenger. To achieve this, it is pertinent to identify the likely potential threats to the autonomous vehicle and then create apposite security solutions to mitigate threats. In this paper, we develop a model for the communication infrastructure to envisage likely threats to the autonomous vehicle and their origin. Then we identified the threats, conduct a risk analysis study to classify risks so as to understand the degree of seriousness of a particular threat and we proposed countermeasures for identify threats using the TVRA methodology by ETSI [5]. \\
The rest of this paper is structured as follows: section II presents the communication infrastructure of the autonomous vehicle. In section III, we present our risk analysis study; we introduce the security environments, security objectives, identified threats, classify risks and propose countermeasures. Section IV concludes this paper. 

\section{Model of Communication Infrastructure}
We present a model of the communication infrastructure for autonomous vehicle in this section. The model encompasses all potential interactions between the vehicle and other connecting entities. \\
P. Kleberger [6] described a communication infrastructure for a connected car. They divided their model into 2 distinct domains; the managed infrastructure and the vehicular communication. The managed infrastructure shows all networks the autonomous vehicle is likely to be connected to while the vehicle communication shows the communication links between the vehicle and other entities. Their model describes the likely connections of an autonomous vehicle therefore we adapt to their model and include only specific connections. Fig.1. describes our communication infrastructure model. \\
\begin{figure*}[t]
\centering
\includegraphics[width=0.6\textwidth]{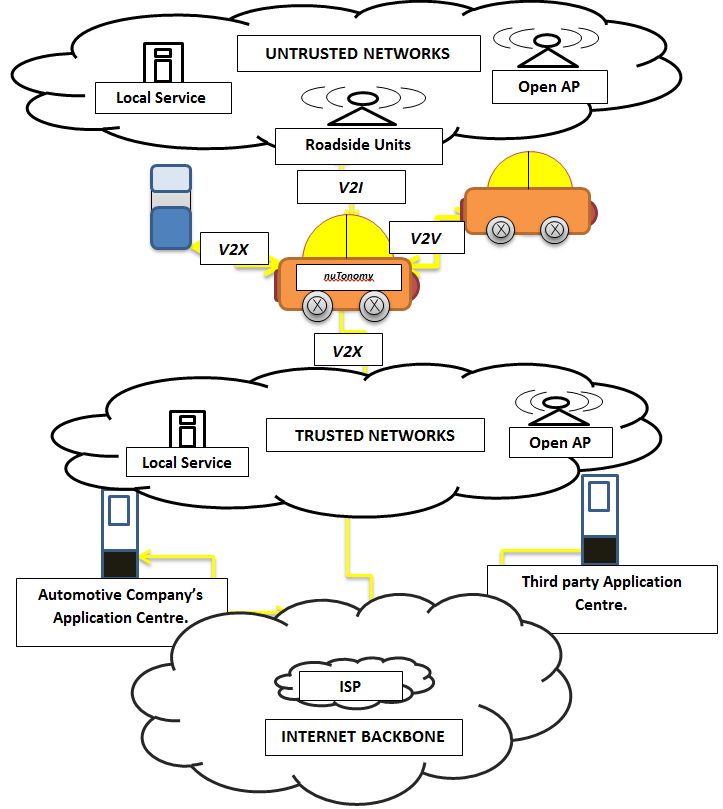}
\caption{Model of the Communication Infrastructure}
%\label{fig_env}
\end{figure*}
In our model, the managed infrastructure includes all networks the vehicle will be connected to and other ancillary services that can be provided for the autonomous vehicle. Vehicle communications shows all form of connections the vehicle will be exposed to using the DSRC wireless communication technology [7]. For example, connection to other vehicles (v2v), to roadside infrastructures (v2i) and other connecting technologies (v2x) like connecting the vehicle to the network of a diagnostic repair shop. 
\section{Risk Analysis}
Having described the model of communication infrastructure for the autonomous vehicles, we analyse the threats and vulnerabilities due to automation of critical vehicle functions and in the communication exchanges between entities in the communication infrastructure model using the TVRA methodology developed by ETSI [5]. Most precisely, we focus on the basic interactions of the autonomous vehicle as it travels; the interactions between the vehicle and other vehicles, the interactions between the vehicle and the road side units and with the internet domain.
\subsection{TVRA Methodology}
 TVRA uses the product estimation of the likelihood of attack and its impact to identify risk due to a particular threat. We employ a subset of the TVRA methodology in our risk analysis study for the autonomous vehicle. Fig.2. specifies the steps we adopted in this study. Furthermore, we modified the TVRA process to include specific objectives for identified threats.  
\begin{enumerate} 
\item Target of Evaluation (ToE): the target of evaluation includes the description of our security environment which depends on the assets, threat agents and the assumptions made. We identify 3 potential targets of evaluation for the purpose of our risk analysis; the autonomous vehicle, the passenger in the vehicle, and the road side infrastructures (RSI).
\begin{itemize}
\item Assumptions:
1.) There exist a passenger in the vehicle; 2.) the passenger  has at least one connecting device; 3.) communications within the target of evaluation boundaries are considered secured; and 4.) the internet is secured from any form of direct attack [5].\\

\begin{figure}[h]
%\centering
\includegraphics[width=0.4\textwidth]{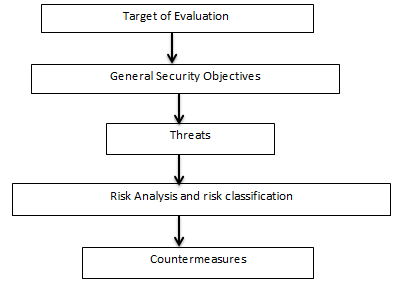}
\caption{TVRA Methodology}
%\label{fig_env}
\end{figure}
\item System assets: From the communication architecture, we categorise the system assets as: physical, logical and human. The physical assets include but not limited to the autonomous vehicle, human machine interface (HMI), road side units (RSU), on-board sensor monitor, and the in-vehicle passenger’s connecting devices. Logical assets include: communication protocols, sensor data, vehicle system controls and ancillary applications such as the applications from the automotive company [2, 5]. The human asset considered in this work is the in-vehicle passenger. While no known work includes this consideration, we argue that human assets could be exploited and used as a means to attack other valid communicating entities. 
\item Threat agents:  we introduce a threat reference model to make obvious the source of threat and identify threat agents. The letters A, B and E in Fig.3 denote access interfaces. These interfaces are points by which an adversary can gain entry into the security environment and compromise a functional entity. The reference point $A_i$ means that in the vehicle, there exist a passenger(s) with connecting devices ranging from (i=1 to n) where n represents the total number of connecting devices the passenger has in the vehicle. The threat agents could be internal or external adversaries who seek an opportunity to compromise the communication infrastructure via a valid ToE. 
\begin{figure}[h]
%\centering
\includegraphics[width=0.50\textwidth]{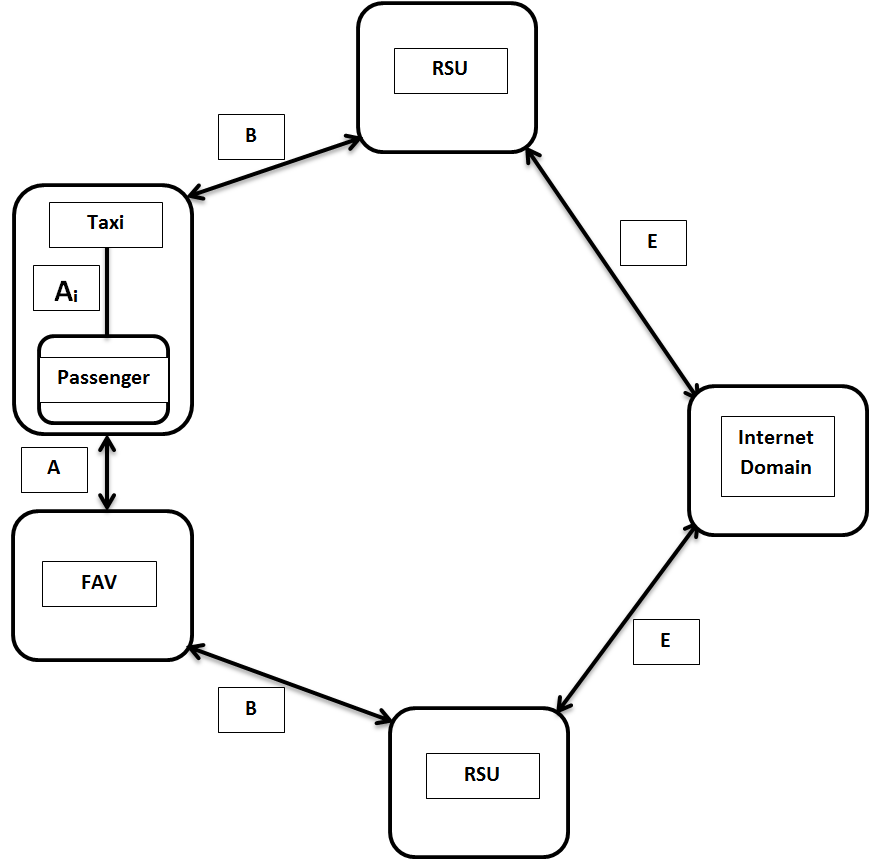}
\caption{Threat Reference Model}
%\label{fig_env}
\end{figure}
\end{itemize}
\item Security Objectives
We classify the security objectives as general and specific. General objectives are defined to specify what security measures needs to exist to assure security of the evaluated entities while the specific objectives are defined for specific threats identified so as to be able to obtain a counter measure that will mitigate threats to the autonomous and connecting system. Specific objectives will be listed in tandem with identified threats. \\
\begin{itemize} 
\item General objectives: 1.) communication between vehicle and other communication entities in the security environment should be secured; and 2.) during major operating processes (booting and storage), the ToE platforms should be secured [2].\end{itemize} 
To achieve the objectives, we consider the following security services to allow for a secure communication [2, 5]; 
\begin{itemize} \item Confidentiality: This requires only legitimate access to broadcasted messages. \item Integrity: This requires that message exchange among authorized entities cannot be altered. \item Availability: This requires that at all time, authorized entities should never be denied access to requisite services. \item Privacy: This requires that the personal information of all human assets are protected from unauthorized access. \item Authentication: This requires that communicating entities can be confirmed as authentic. \item Authorization: This requires that only entities with necessary privileges have access to certain information. \item Plausibility: This ensures the veracity of data. \item Non-repudiation: This requires the implementation of an audit trail to keep track of executed transactions. \end{itemize}
\item Threats 
We categorize the threats identified in this study as data, logical and human threats. Data threats are directed towards the identification of the asset and its location in a ToE environment. Logical threats are a compromise to the logical assets and directed towards the exploitation of a device, item or application. Human threats are threats carried out by authorized entities or directed to them. Table I summarizes the identified threats.  
\begin{table*}[h]
  \centering
  \begin{center}
  \caption{Threats}
  \vspace{0ex}
  \begin{adjustbox}{width=1.0\textwidth}
    \begin{tabular}{| p{2.0cm}| p{3.0cm}| p{5.0cm}| p{2.0cm}| p{6.0cm}|} \hline
    \textbf{Category of Threat} & \textbf{Type of Threat}  & \textbf{Action of Attacker} & \textbf{Attack Interface}   & \textbf{Specific Objectives}   \\ \hline
     Data Threat  & Eavesdropping and Traffic analysis. & 1) Alter data to reveal identify vehicle.  2) Alter data to reveal identity of passenger.  &  A, B & Deny access of legitimate traffic to unauthorized entities.\\ \cline{2-5} 
    %\vspace{1ex}
    & Location Tracking & Alter data to reveal location of vehicle  &  A, B & Ensure security of network traffic to deny access to unauthorized entities.\\ \cline{2-5} 
    & Message Injection  & 1) Generate and sends false information on a message. &  A, B & Deny access to unauthorized entities. \\  \cline{2-5}
  	   & Sensor spoofing  & 1) Alter data to change the location of the vehicle.\ 2) Malicious use of sensors to generate faulty data &  A, B & Deny access to unauthorized entities \ and secure communication channels. \\ \cline{2-5}
  	   & Replay  & Resend old or expired message to obtain network access. &  A, B & Verification of incoming messages.\\  \cline{2-5}
  	   & Timing Attack  & Delay sending safety and time critical message. &  A, B, B on behalf of E & Ensure data  cannot \ be altered during transmission.\\  \hline
  	 Logical Threat  & Illusion Attack. & Alter sensors to deliver fake messages &  A, B & Validation of sensor data.\\ \cline{2-5}  
  	  & Masquerade. & 1) Obtain sensitive information by acting as a legitimate entity 2) Introduce fake messages into the network by posing as an authorized entity. &  A, B & 1) Enforce strict security policies on authentication. 2) For message reliability, ensure its integrity.\\ \cline{2-5}
  	  & Denial of service. & 1) Flooding: Overload the network with spurious information. 2) Spamming: Increase transmission latency and use up network bandwidth by sending high volume of messages intentionally. 3)Blackhole: Drop re-routing and misdirecting messages in the network. 4) Jamming: Cause message loss by intentionally creating interference.  &  A, B, B on behalf of E & Deny illegitimate network entry.\\ \cline{2-5}
  	   & RF-Fingerprinting. & identify source and location of radio transmission &  A, B & Ensure that it is very difficult to identify a functional entity.\\ \cline{2-5}  \
  	   & Sybil Attack & Deliver fake messages by claiming to be multiple nodes with different identities &  A, B  & Validation of sensor data \\ \cline{2-5}
  	   & Message suppression. & Information loss due to intentional message deletion &  A, B & Allow access to authorized entities alone and implement an audit trail to keep track of actions.\\ \cline{2-5}
  	   & Malware. & Flash ECU to execute malicious software &  A, B & Secure update of firmware patches.\\ \cline{2-5}
  	  % & File Modification. & Alteration of config. files &  A, B & Network exploitation & Ensure access of legitimate resources to specific authorized entities & Access control, Non-repudiation and confidentiality. \\ \cline{2-7}
  	   & Repudiation. &  Deny received or sent transmissions &  A, B & Maintain an audit trail to keep track of executed actions.\\ \hline
  	   Human Threat  & Social Engineering Attacks (Phishing). & Compromise mobile devices of in-vehicle passenger & {$A_i$}, A, B and B on behalf of K & Detection and prevention of misbehaving users.\\ \cline{2-5}
  	   & Intrusion attacks. & Installation of a compromised device into the vehicle by the in-vehicle passenger & {$A_i$}, A & Security policies/standards for external device usage in-vehicle.\\ \hline	   
    \end{tabular}%
    \end{adjustbox}
  \label{tab:addlabel}%
\end{center}
\end{table*}%

\item Risk analysis 
We obtain the risks for identified threats using an estimation value of likelihood that an attack will occur and the impact of the threats to the system. 
\begin{align}
\text{Risk} &= \text{Likelihood * Impact}
\end{align}
To obtain the likelihood of an attack, we give an attack potential based on time it will take to plan and launch the attack, the expertise of the attacker, knowledge, opportunity and the equipment needed to conduct the attack. The derived value is then mapped to the TVRA vulnerability rating to determine likelihood of occurrence [8, 9]. \\
\begin{table}[h]
  \centering
  \begin{center}
  \caption{Vulnerability rating and likelihood}
  %\vspace*{1ex}
  \begin{adjustbox}{width=1.0\linewidth}
    \begin{tabular}{| p{2cm}| p{3cm}| p{1cm}|} \hline
    \textbf{Vulnerability Rating} & \textbf{Likelihood}  & \textbf{Values} \\ \hline
    %\vspace{1ex}
    Beyond High  & Unlikely &  1\\ \cline{0-0}
    High &  &  \\ \hline
    Moderate  & Possible & 2 \\ \hline
    Basic  & Likely &  3\\ \cline{0-0}
    No-rating &  &  \\ \hline
    \end{tabular}%
    \end{adjustbox}
  \label{tab:addlabel}%
\end{center}
\end{table}%
The impact reveals the likely consequences to the ToE. Impact is classified as high (3): threat has serious consequences. Medium (2): threat cannot be neglected; and Low (1): likely damage due to threat is low [3]. Risks are classified as minor (1,2,3)  meaning the attack is unlikely, major (4) meaning the attack is likely but consequence is likely not fatal and critical (6,9) meaning the risk is should be treated as urgent  and appropriate countermeasures must be developed for them [2, 9]. Fig. 4 details the risk classification of the identified threats. \\
\begin{figure*}[h]
\centering
\includegraphics[width=0.78\textwidth]{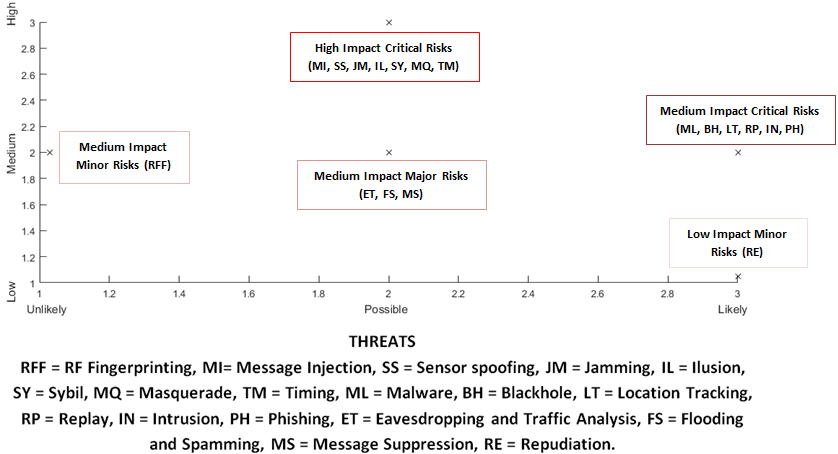}
\caption{Risk Analysis}
%\label{fig_env}
\end{figure*}

In Fig. 4, all high impact threats have been classified as critical risks. These threats are also means by which other attacks could be mounted on the exposed ToE interfaces. For example,  Illusion attack gives an attacker an advantage on the road. The attacker suppresses, alters messages and also communicates false sensor readings to deceive neighbouring vehicles.  With sensor spoofing an attacker can cause a holistic dysfunction of the communication infrastructure. GPS time for example is used for time stamping and synchronizing messages. However with GPS spoofing, an attacker can impact on message timing by means of replay attack [2]. Sybil attack allows for the consumption of network bandwidth thereby leading to a denial of service attack. The serious impact of these attacks calls for the development of urgent and priority countermeasures to mitigate these threats. \\
Reference points A, B represent a significant entry point for malicious users. High impact threats and major risks also emanate from reference points A, B therefore, we conclude that the security of devices in the ToE exposed to reference points A, B would guarantee the overall security of the autonomous vehicle.  \\
Reference points $A_i$ and A represents a significant entry point for malicious users. Threats from these reference points are critical. This shows that a successful exploitation by a human asset has serious consequences on the communication infrastructure. Reference point $A_i$ represents the attack interface for the in-vehicle passenger who via spoofing and intrusion attacks compromise or could be used to compromise the communication infrastructure.
\item Countermeasure: 
The countermeasures to specific threats have been identified in Table III. \\
\begin{table}[h]
  \centering
  \begin{center}
  \textbf{\caption{Potential countermeasures to threats}}
  \vspace{1ex}
  \begin{adjustbox}{width=1.18\linewidth}
    \begin{tabular}{| p{3.0cm}| p{1.5cm}| p{7.0cm}|} \hline
  \textbf{Threats}  & \textbf{Risk} & \textbf{Countermeasure} \\ \hline
    %\vspace{1ex}
	Eavesdropping and Traffic analysis & Major & 1) Secure transmission of personal and private data via encryption. 2) Implement pseudonyms so that driver, passenger or vehicle are uniquely unidentifiable. \\ \hline
	Message Injection & Major & Thorough validation of input.  \\ \hline
	Spoofing & Critical & 1) Packet inspection. 2) Encryption of transmission medium.   \\ \hline
	Replay & Critical &  The use of a pseudo-random session token and time-stamp by means of a secured time synchronization protocol.    \\ \hline
	Flooding and Spamming & Major & 1) Implement an intrusion detection system. 2) Use packet filters to block unwanted traffic. \\ \hline
	Blackhole & Critical & 1) Implement an intrusion detection and prevention system. 2) Use trust based routing protocols. \\ \hline
	Jamming & Critical & 1) Channel switching 2) Track jammers and block them using a jamming detection device.  \\ \hline
	Illusion & Critical & Implement a plausibility validation mechanism for sensor data verification  \\ \hline
	Phishing & Critical & User identity verification via a 2-step authentication procedure  \\ \hline
	Network Intrusion & Critical & Use intrusion detection and prevention mechanisms  \\ \hline
	 Sybil  & Critical & 1) Implement a position verification protocol 2) Use digital signatures and certificate authorities and ensure each network device has one digital signature and certificate authority  \\ \hline
	 Message suppression & Major & Ensure that communication channels are encrypted as well as data transmitted. \\ \hline
	 Malware & Critical & a Use intrusion detection and prevention mechanisms (packet filters)  \\ \hline
	  Masquerade & Major & 1) Ensure that vehicles have unique and verifiable public keys (scenario I). 2) Plausibility checks on incoming messages. \\ \hline
	  Timing & Critical & Verification of data integrity via cryptographic means to secure hardware, protect and store data.  \\ \hline
	  Backdoor & Critical & Use audit-trails to keep track of actions. 2.) if backdoor is on a hardware device requiring physical access, implement physical access control and non-repudiation policies.   \\ \hline
	  Repudiation & Low &  Auditing and Logging.   \\ \hline

   \end{tabular}%
    \end{adjustbox}
  \label{tab:addlabel}%
\end{center}
\end{table}%
The influence of human assets over the communication infrastructure must be critically considered and appropriate countermeasures built around intrusion detection, prevention and encryption mechanisms as well as strong security policies must be ensured to prevent and mitigate threats. \\
\end{enumerate}

\section{Conclusion}
In this paper, we have described a communication model for autonomous vehicles and identified threats to the security of the full autonomous vehicle communication infrastructure. We conducted a risk analysis study using the TVRA methodology developed by ETSI and proposed countermeasures for identified threats. In analysing the risk to the communication infrastructure, we considered human assets as a means of exploitation and discovered that the impacts of the threat from humans are critical to the security of the communication infrastructure. The classification of risks for threats identified represents an upper-bound risk to the autonomous vehicle. The challenge now is  expedite development of apposite security mechanisms to mitigate threats to facilitate the adoption of the autonomous vehicle technology. \\

%\end{enumerate}

\end{document}